\documentclass[a4paper]{jpconf}
\usepackage{graphicx}
\begin{document}

\title{Topological signature in the NEXT high pressure xenon TPC}
\author{Paola Ferrario for the NEXT Collaboration}
\address{Instituto de F\'isica Corpuscular (IFIC), CSIC \& Universitat de Val\`encia,
Calle Catedr\'atico Jos\'e Beltr\'an, 2, 46980 Paterna, Valencia, Spain}

\ead{paola.ferrario@ific.uv.es}

\begin{abstract}
The NEXT experiment aims to observe the neutrinoless double
beta decay of \ensuremath{{}^{136}\rm Xe} in a high-pressure xenon gas TPC using
electroluminescence to amplify the signal from ionization. One of the main advantages of this technology is the
possibility to use the topology of events with energies close to $Q_{\beta\beta}$ as an extra tool to reject background.
In these proceedings we show with data from prototypes that an extra background rejection factor of 24.3 $\pm$ 1.4 (stat.)\% can be achieved, while maintaining an efficiency of 66.7 $\pm$ 1.\% for signal events. The performance expected in NEW, the next stage of the experiment, is to improve to $12.9\% \pm 0.6\%$ background acceptance for $66.9\% \pm 0.6\%$ signal efficiency. 
\end{abstract}

\section{The NEXT experiment}

NEXT (Neutrino Experiment with a Xenon TPC) searches for neutrinoless double beta decay in a high pressure xenon TPC at the Canfranc Underground Laboratory (LSC, from its initials in Spanish), in the Spanish Pyrenees. A charged particle propagating in the gas deposits its energy through both scintillation and ionization of the gas molecules. The scintillation light 
is registered by photomultipliers (PMTs) on the cathode side and gives the starting time of the event. The ionization electrons are drifted by an electric field all the way through the drift region until they enter a region of moderately higher field where they are accelerated and secondary scintillation (but not ionization) occurs. This process, called electroluminescence (EL), results in an amplification of the signal. 
The PMTs detect the EL light, giving a precise measurement of the energy of the event. On the anode side, the distribution of the EL light every microsecond on a grid of silicon photomultipliers (SiPMs) placed behind the EL area is a 2D picture of the track at a given position along the axis. Knowing the starting time of the event, the absolute position along the TPC axis can also be reconstructed.

The signal of a neutrinoless double beta decay is a peak in the kinetic energy spectrum of the outcoming electrons ($Q_{\beta\beta}$).
For this reason an experiment must be optimised simultaneously for energy resolution and the rejection of background events with energies similar to $Q_{\beta\beta}$.  While electrons and muons coming from outside can be efficiently vetoed with fiducial cuts, high energy gammas from environmental radioactivity entering the active volume of the detector are the main background in NEXT. When gammas interact with xenon gas, they can produce photoelectric and Compton electrons, at energies very similar to Q$_{\beta\beta}$, which can mimic the signal \cite{Martin-Albo:2015rhw, neus}.
 
NEXT has proven an excellent energy resolution (0.74\% FWHM extrapolated to the Q$_{\beta\beta}$ of \ensuremath{{}^{136}\rm Xe}, i.e., 2.458 MeV)  \cite{Lorca:2014sra} and topological signature for background rejection \cite{Ferrario:2015kta} in prototypes. 

\section{The topological signature}

Electrons moving through xenon gas lose energy at an approximately fixed rate until they become non-relativistic. At the end of the trajectory the 1/$v^2$ rise of the energy loss (where $v$ is the speed of the particle) leads to an energy `blob', i.e., a high energy deposition in a compact region. This feature can be used to distinguish background single-electrons (one `blob' only at one extreme) from signal double-electrons (a single track with two `blobs' at the end-points). 

\section{Track reconstruction and topological cut}\label{sec:topoCut}

The current track reconstruction is based on the analysis of the charge detected in each time bin by the SiPMs. A search for 2D hits and a subsequent voxelization of the whole space is performed, and a Breadth First Search (BFS) algorithm is used to connect the voxels to form tracks. The algorithm sorts the voxels into tracks with a criterium of connectivity, which considers two voxels as connected if their centres are closer than a maximum distance. The algorithm also finds the end-points of a track as the voxels with maximum distance along the track.

A `blob' candidate is defined as a group of voxels which fall within a set radius from a track end-point. The energies of the `blob' candidates have similar values for signal events, while the `blob' candidate with less energy has a much lower energy for background. Therefore, a threshold can be set on the energy of the lower energy `blob' candidate to select signal events.


\section{Results in NEXT-DEMO data}

\begin{figure}[htb]
\begin{center}
\includegraphics[width=0.4\textwidth]{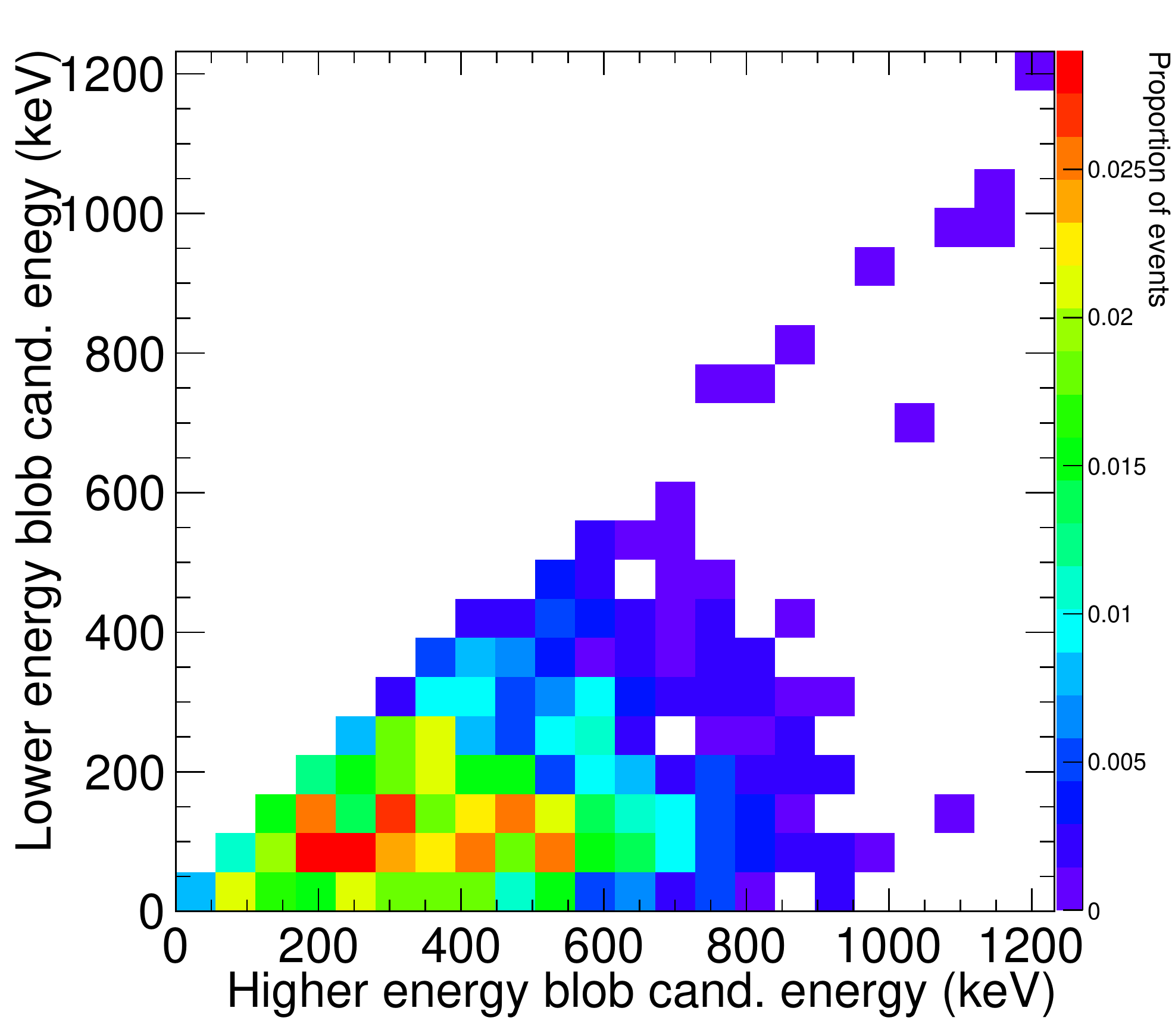} \hspace{3.pc}
\includegraphics[width=0.4\textwidth]{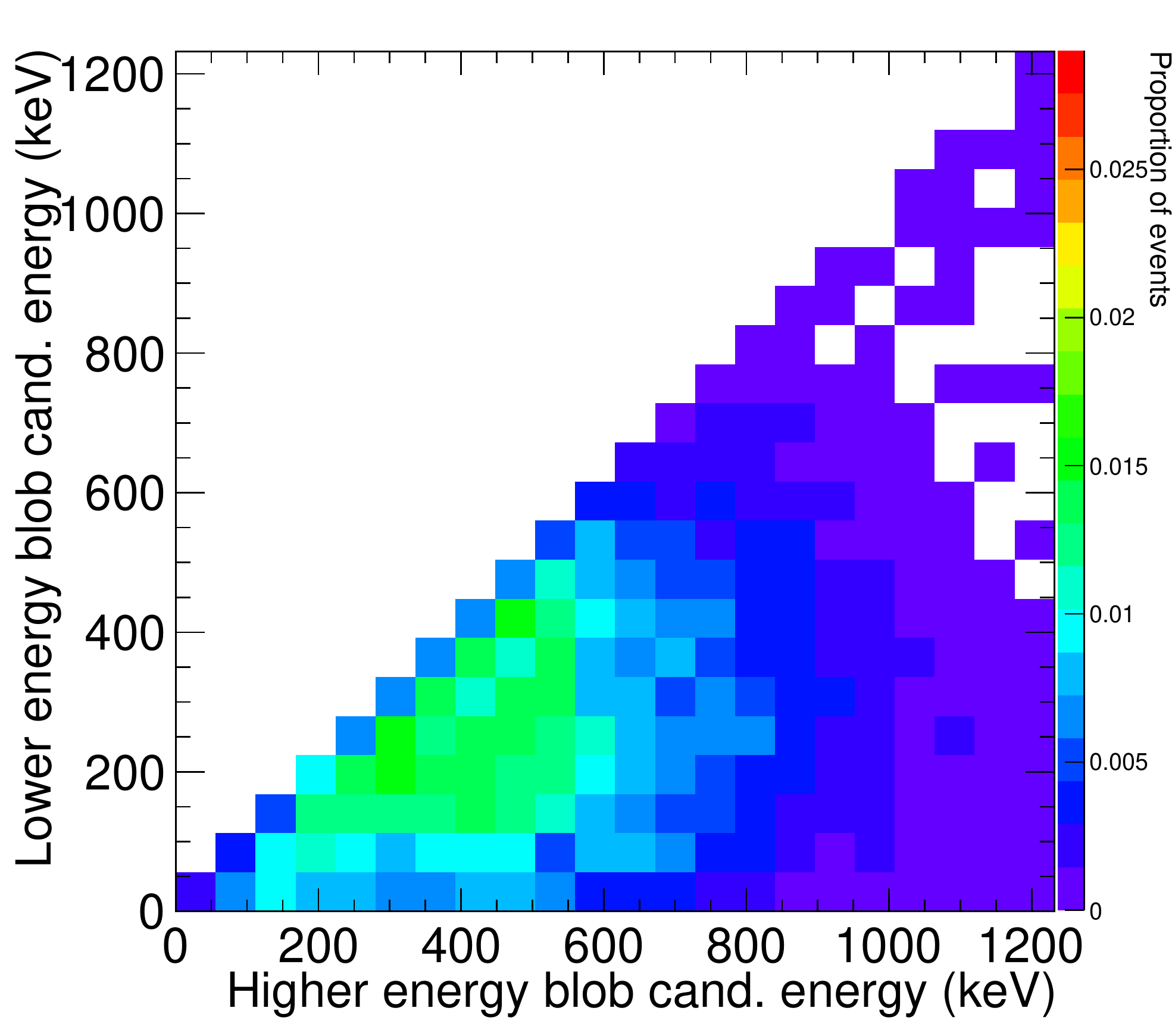}
\caption{Energy distribution at the end-points of the tracks coming from \ensuremath{{}^{22}\rm Na}  decay (left) and those coming from the \ensuremath{{}^{228}\rm Th} decay (right) for 2 cm radius `blob' candidates.}
\label{fig:2blobs}
\end{center}
\end{figure}

NEXT-DEMO was a 1--kg prototype, built and operated at IFIC, Valencia. With this prototype, the power of the topological cut has been demonstrated  using radioactive sources: \ensuremath{{}^{22}\rm Na} provides high energy electrons and \ensuremath{{}^{228}\rm Th} electron-positron pair production, to mimic background and signal respectively. The topological cut described in section \ref{sec:topoCut} was applied as last stage of the analysis, with an optimized threshold of 210 keV. Figure \ref{fig:2blobs} shows the energy distribution of the `blob' candidates for signal-like and background-like events.
A signal efficiency of $66.7\% \pm 0.9\%$ and a background acceptance of $24.3\% \pm 1.3\%$ is found, in good agreement with Monte Carlo simulations.

\section{Expected performance in NEW}


\begin{figure}[h]
\includegraphics[width=20pc]{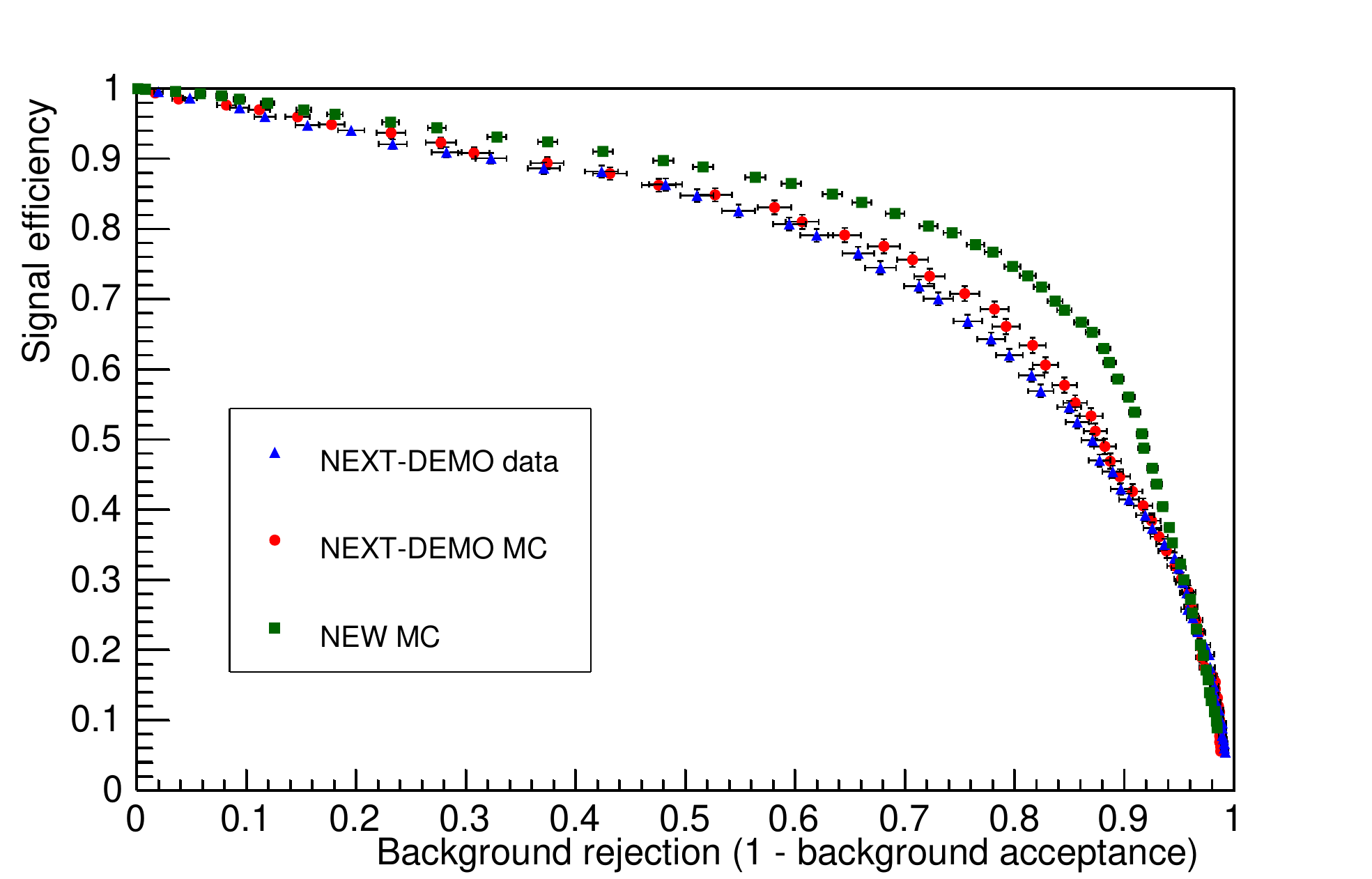}\hspace{2pc}%
\begin{minipage}[b]{14pc}\caption{\label{fig:dataMC} Signal efficiency as a function of background rejection (proportion of background events removed from the sample by the two--`blob' cut) varying the required minimum energy of the lower energy `blob' candidate. DEMO data, DEMO Monte Carlo and NEW Monte Carlo are shown for comparison.}
\end{minipage}
\end{figure}

A first stage of the experiment, NEW \cite{miquel}, with $\sim$10 kg of xenon, is being commissioned at LSC. NEW will be used for background and two-neutrino double beta decay measurements, as well as to prove energy resolution and the power of topological rejection at energies close to $Q_{\beta\beta}$. Simulations indicate a significant improvement of the topological rejection, due to the larger volume of the detector. First Monte Carlo studies point to $66.9\% \pm 0.6\%$ signal efficiency for $12.9\% \pm 0.6\%$ background acceptance for the same analysis as in NEXT--DEMO at a pressure of 10 bar. Figure \ref{fig:dataMC} shows a comparison of the signal efficiency versus background rejection between data and Monte Carlo in NEXT--DEMO and the predictions for NEW.

\section{Future improvements}

New reconstruction approaches are being investigated \cite{Simon:2016yso,Renner:2016lvq}. The Maximum Likelihood Expectation Minimization method tries to solve the inverse problem of finding a set of energy depositions in the chamber, given the sensor response to the EL light. Given a statistical model that describes the forward problem, it provides estimates for its parameters, maximizing the likelihood of the model, given any outcome. We are also exploring the power of deep neural networks, which could be used for reconstruction and classification of events as signal or background, exploiting all possible features in the image.


\ack
The NEXT Collaboration acknowledges support from:
the European Research Council (ERC) under the Advanced Grant 339787-NEXT;
the Ministerio de Econom\'ia y Competitividad of Spain and FEDER under grants CONSOLIDER-Ingenio
2010 CSD2008-0037 (CUP), FIS2014-53371-C04 and the Severo Ochoa Program
SEV-2014-0398; GVA under grant PROMETEO/2016/120. Fermilab is operated by Fermi Research Alliance, LLC under Contract No. DE-AC02-07CH11359 with the United States Department of Energy.

\bibliographystyle{iopart-num}

\section*{References}

\bibliography{biblio}


\end{document}